\documentclass[useAMS,usenatbib]{mn2e} \usepackage{graphicx}

\title[G12.889+0.489 hydroxyl variability]{Variability monitoring of the hydroxyl maser emission in G12.889+0.489}

\author[Green et al.]{J. A. Green\thanks{E-mail:james.green@csiro.au}, J. L. Caswell, M.A. Voronkov and N. M. McClure-Griffiths\\
CSIRO Astronomy and Space Science, Australia Telescope
National Facility, PO Box 76, Epping, NSW 2121, Australia\\}
\date{Accepted 2012 July 3. Received 2012 July 2; in original form 2012 May 24}

\pagerange{\pageref{firstpage}--\pageref{lastpage}} \pubyear{XXXX}
\begin{document} \maketitle

\label{firstpage}

\begin{abstract}
Through a series of observations with the Australia Telescope Compact Array we have monitored the variability of ground-state hydroxyl maser emission from G12.889+0.489 in all four Stokes polarisation products.  These observations were motivated by the known periodicity in the associated 6.7-GHz methanol maser emission. A total of 27 epochs of observations were made over 16 months. No emission was seen from either the 1612 or 1720 MHz satellite line transitions (to a typical five sigma upper limit of 0.2 Jy). The peak flux densities of the 1665 and 1667 MHz emission were observed to vary at a level of $\sim$20\% (with the exception of one epoch which dropped by $\le$40\%). There was no distinct flaring activity at any epoch, but there was a weak indication of periodic variability, with a period and phase of minimum emission similar to that of methanol. 
There is no significant variation in the polarised properties of the hydroxyl, with Stokes $Q$ and $U$ flux densities varying in accord with the Stokes $I$ intensity (linear polarisation, $P$, varying by $\le$20\%) and the right and left circularly polarised components varying by $\le$33\% at 1665-MHz and $\le$38\% at 1667-MHz. These observations are the first monitoring observations of the hydroxyl maser emission from G12.889+0.489.
\end{abstract}


\section{Introduction}
A site of hydroxyl maser emission, G12.889+0.489, initially found in a search towards IRAS 18089$-$1732 \citep{cohen88}, was found to also be a prominent methanol maser at 6.7 GHz  \citep{menten91}.  Variability of the methanol was identified by \citet{caswell95d} and led to the extensive monitoring of \citet{goedhart04} and the discovery of periodicity \citep{goedhart09}.  6.7-GHz methanol masers exclusively trace the formation of high-mass stars and are hence closely studied to gain insight into the largely unknown processes of high mass star formation. The discovery of periodicity in the methanol maser emission indicates a periodic variation in the inputs of the maser emission process which has significant implications both for the mechanism of maser emission and the nature of high-mass star formation, indicating periodic processes such as the interaction of winds from binary pre-main sequence high mass stars \citep[e.g.][]{walt09}.

\citet{goedhart09} found intensity variations in the methanol emission with a period of nearly 30 days that appeared stable, although variations from cycle to cycle in the peak amplitude of the flare were apparent. The flaring features peaked anywhere within an 11-day window, but the phase of the minima was stable. An amplitude variation with the same period is present in features at different velocities and at both 6.7 GHz and 12.2-GHz transitions. Delays of up to six days were found between individual 6.7-GHz features and a one day delay was found between the 12.2-GHz and 6.7-GHz flares at the same velocities. 

Hydroxyl observations since the discovery of the maser by \citet{cohen88} have included the positioning observations of \citet{caswell98} and \citet{argon00}, and a detailed polarization study by \citet{szymczak09}. Apparent variations in total intensity exceed 20\%, but intervals between observations were typically longer than one year and precise comparison is not possible with the significantly different instrumentation. Observations with the Parkes radio telescope (Caswell et al. in prep.) measured the hydroxyl emission in 2004 and 2005, finding peak flux densities of 8 Jy at 1665 MHz and 1.8 Jy at 1667 MHz. The Parkes spectra are in agreement at the two epochs to within 10\% and similar values were obtained by \citet{szymczak09} in observations made in 2003. Although the amplitudes have shown some variation, the velocities of the features are consistent over all previous observations.

The majority ($\sim$80\%) of hydroxyl masers in regions of high-mass star formation have associated methanol maser emission \citep{caswell98}, with many sites exhibiting a close spatial coincidence, implying a common masing gas and pumping source. Models of maser pumping suggest that both species are pumped by infrared emission from dust surrounding the high mass pre-main sequence object \citep[e.g.][]{moore88, cragg05, gray07}. Hence there is an expectation that any variability in emission of one species will correlate with variability in the other. Early evidence for hydroxyl maser variability  \citep[e.g.][]{robinson70} was followed by several more focused variability studies  \citep[e.g.][]{sullivan76, clegg91}, but these showed no clear periodicity of hydroxyl masers in star forming regions. More recent investigations of possible common flaring behaviour between the methanol and the ground-state hydroxyl masers has been inconclusive \citep[e.g.][]{macLeod96}, but correlated variability of excited-state hydroxyl with methanol \citep{almarzouk12} and  formaldehyde with methanol \citep{araya10} has been found recently.  With methanol emission from G12.889+0.489 established to be periodic in nature, we were motivated to initiate monitoring observations of the associated hydroxyl emission.

\section{Observations}
Observations were made with the Australia Telescope Compact Array (ATCA), commencing with a full synthesis image on  2010 May 20 consisting of four cuts of 10 minutes spread over eight hours. This was followed by 26 epochs over 1.5 years with typical integration times of 40 minutes on source. These observations were divided into two groups, the first with one to two observations per week for $\sim$10 weeks, the second with one observation approximately every 24 hours for seven days. Observations were centred at 1612.2310 MHz, 1665.4018, 1667.3590 MHz and 1720.5300 MHz, corrected for the motion of the local standard of rest. The ATCA calibrator 1830$-$210 was used for phase calibration for the first group of observations and ATCA calibrator 1829$-$207 for the second group (with positional uncertainties of $<$0.15$''$ and $<$0.01$''$ respectively). Both groups used the primary ATCA flux density calibrator PKS 1934$-$638, with flux densities bootstrapped to this (with a resultant systematic uncertainty of $<$2\%). All observations were made with the new Compact Array Broad Band backend \citep{wilson11}, adopting the CFB 1M-0.5k mode (2048 channels over 1 MHz giving 0.5 KHz channel spacings), obtaining all four polarisation products. The velocity channel separations were 0.091, 0.088, 0.088 and 0.085\,km\,s$^{-1}$ for the 1612, 1665, 1667 and 1720 MHz transitions respectively. The full listing of the observations and noise levels is given in Table\,\ref{dates}. Antenna 4 was not used for epochs seven to 18 due to the upgrade of the low frequency (1$-$3 GHz) receiver. 
Observations were made within the LST range 14:30 to 22:00 hrs and predominantly in the 6-km array configurations (6B and 6C), but eight epochs were observed with compact configurations (see Table\,\ref{dates} for details). There were no systematic variations in flux density for the observations with the short baseline (compact) array configurations compared with the long baseline (6-km) configurations.
The data were reduced and processed with the {\sc miriad} software package using standard techniques \citep{sault04}. Following the positioning observation, spectral profiles were obtained with the task {\sc uvspec} and a time series of peak flux density of maser features was determined through fitting a Gaussian in the spectral domain with $\chi$$^{2}$ minimisation for each epoch. 

\begin{figure}
\begin{center}
\renewcommand{\baselinestretch}{1.1}
\includegraphics[width=8.0cm]{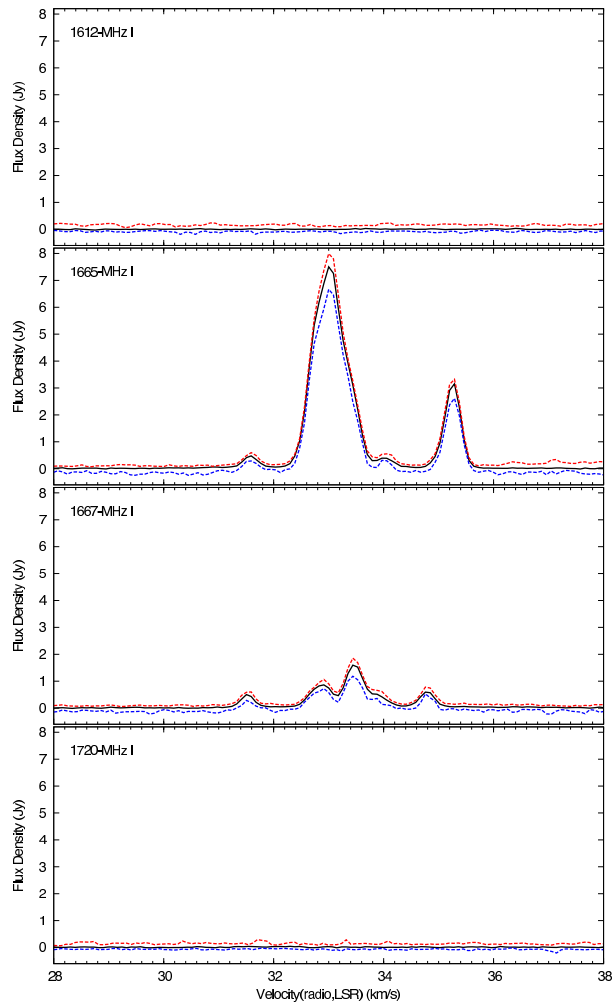} 
\caption{\small Range in flux density variation in Stokes $I$ for G12.889+0.489 at each ground-state hydroxyl transition. The solid line is the averaged spectrum, constructed from the median value from all epochs for each spectral channel. The two dashed lines show the net extreme values (having excluded the most extreme noise-biased values).}
\label{rangefig1}
\end{center}
\end{figure}

\begin{table*} \centering
 \caption{\small Details of observations. ` - ' mark observations where data were unusable. The first observation, denoted {\it `imaging'} was a synthesis imaging observation as discussed in the text.} 
\begin{tabular}{clccccccc}
\hline
\multicolumn{1}{c}{epoch} & \multicolumn{1}{c}{Gregorian} & \multicolumn{1}{c}{Modified} & \multicolumn{1}{c}{Array} & \multicolumn{1}{c}{Integration}& \multicolumn{4}{c}{$\sigma$$_{rms}$}\\
&\multicolumn{1}{c}{Date}&Julian&Config.&Time&1612&1665&1667&1720\\
\multicolumn{1}{c}{} &\multicolumn{1}{c}{} & \multicolumn{1}{c}{ Date} & \multicolumn{1}{c}{} & \multicolumn{1}{c}{ (mins)} & \multicolumn{1}{c}{(mJy)}& \multicolumn{1}{c}{(mJy)}& \multicolumn{1}{c}{(mJy)}& \multicolumn{1}{c}{(mJy)}\\
\hline
{\it imaging} & 2010-05-20 & 55336.5 &6C& 32&57 & 64 & 49 & 64 \\
1 & 2010-05-24 & 55340.5 &6C& 40&44 & 59 & 49 & 56 \\
2 & 2010-06-03 &  55363.5&6C& 40&34 & 35 & 42 & 41 \\
3 & 2010-06-07 &  55367.5&6C& 40&69 & 42 & 39 & 58 \\
4 & 2010-06-10 &  55370.5&6C& 40&32 & 41 & 34 & 49 \\
5 & 2010-06-14 &  55374.5&6C& 40&34 & 34 & 34 & 44 \\
6 & 2010-06-18 &  55378.5&6C& 70&29 & 32 & 33 & 41 \\
7 & 2010-06-27 &  55387.5&6C& 60&38 & 40 & 49 & 42 \\
8 & 2010-06-29 &  55389.5&6C& 40&50 & 51 & 57 & 65 \\
9 & 2010-07-02 &  55392.5&6C& 40& - & 55 & 49 & - \\
10 & 2010-07-04 &  55394.5&6C& 40&45 & 46 & 53 & 51 \\
11 & 2010-07-08 &  55398.5&1.5D& 45&40 & 62 & 62 & 45 \\
12 & 2010-07-15  & 55405.5&EW352& 40&200 & 137 & 141 & 259 \\
13 & 2010-07-18  & 55408.5&EW352& 40&184 & 219 & 131 & 254 \\
14 & 2010-07-21  & 55411.5&EW352&45 &113 & 90 & 91 & 102 \\
15 & 2010-07-27  & 55417.5&H168& 80&52 & 53 & 57 & 57 \\
16 & 2010-07-29  & 55419.5&H168&45 &56 & 61 & - & 68 \\
17 & 2010-08-01  & 55422.5&H168&40 &82 & 90 & 91 & 96 \\
18 & 2010-08-05  & 55426.5&H168& 72&46 & 66 & 55 & - \\
19 & 2011-08-26 &55812.5& 6B&40&33 & 37 & 36 & 36\\ 
20 & 2011-08-27 &55813.5& 6B&40&42 & 37 & 36 & 38\\
21 & 2011-08-29 &55815.5& 6B&40&34 & 39 & 39 & 35\\
22 & 2011-08-30 &55816.5& 6B&40&37 & 41 & 35 & 35\\
23 & 2011-08-31 &55817.5& 6B&40&33 & 35 & 37 & 35\\
24 & 2011-09-01 &55818.5& 6B&40&48 & 39 & 34 & 36\\ 
25 & 2011-09-02 &55819.5& 6B&40&43 & 38 & 43 & 39\\
26 & 2011-09-03 &55820.5& 6B&40&31 & 37 & 34 & 35\\
\hline
\end{tabular} 
\label{dates}
\end{table*}

\section{Results}
\begin{figure*}
\begin{center}
\renewcommand{\baselinestretch}{1.1}
\includegraphics[width=17.5cm]{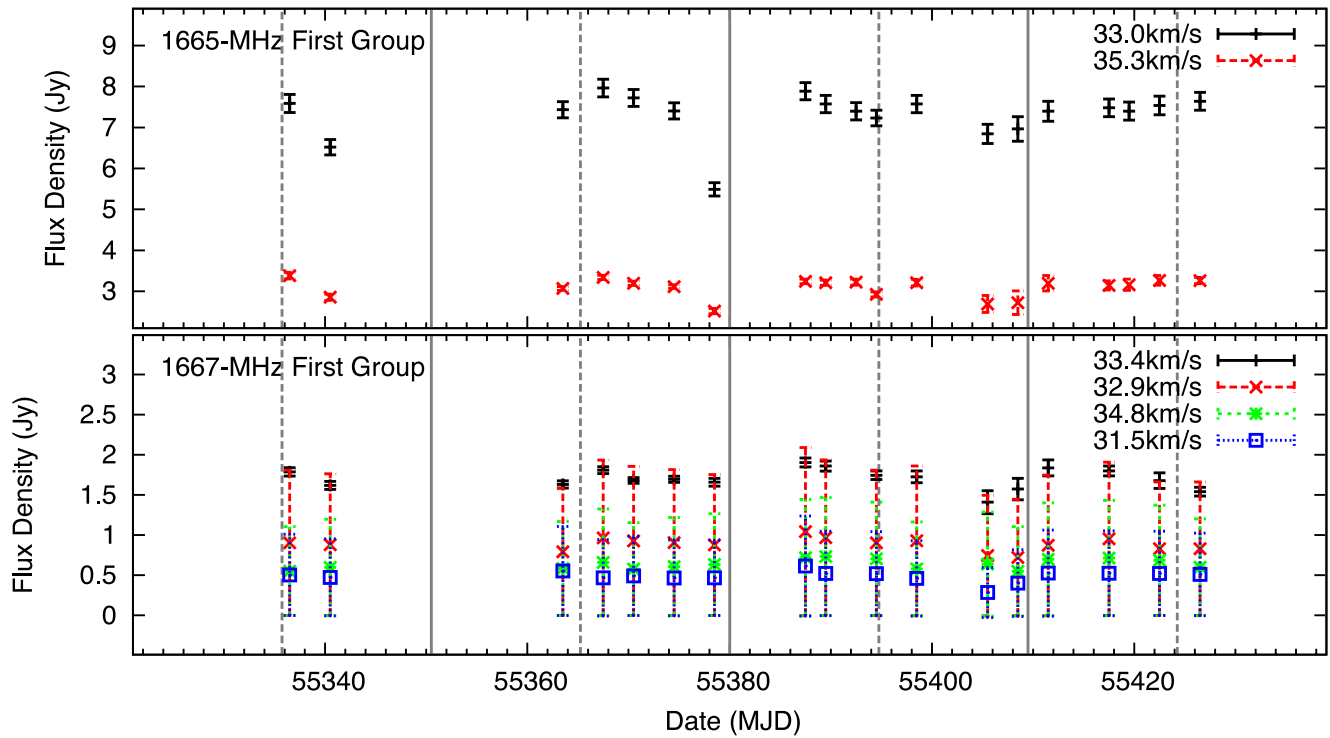} 
\caption{\small Time series for the two brightest features at 1665 MHz and the four brightest features at 1667 MHz in Stokes $I$ for the first group of observations, spanning 90 days. Errors are combination of the rms noise and the errors in the Gaussian fit in the spectral domain. Solid (dashed) grey vertical lines show epochs of extrapolated minima (and half a period offsets) of methanol emission from \citet{goedhart09}.}
\label{timeseriesfigure1}
\end{center}
\end{figure*}

\begin{figure}
\begin{center}
\renewcommand{\baselinestretch}{1.1}
\includegraphics[width=7.5cm]{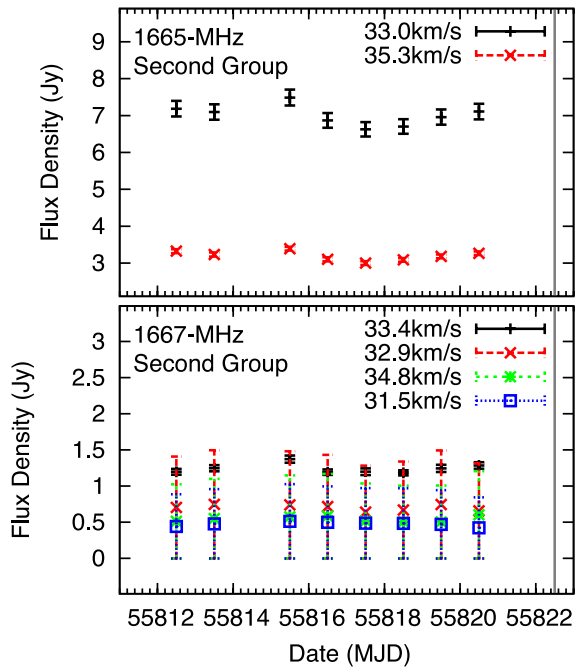} 
\caption{\small Time series for the two brightest features at 1665 MHz and the four brightest features at 1667 MHz in Stokes $I$ for the second group of observations, spanning 8 days. Errors are combination of the rms noise and the errors in the Gaussian fit in the spectral domain. Solid grey vertical line shows  an epoch of the extrapolated minima of methanol emission from \citet{goedhart09}.}
\label{timeseriesfigure2} 
\end{center}
\end{figure}

The position of strongest emission at 1665 and 1667 MHz from the initial (full synthesis imaging) observation was found to be RA(J2000) 18$^{\rm h}$11$^{\rm m}$51.45$^{\rm s}$ and Dec(J2000) $-$17$^{\circ}$31$'$29.7$''$ with a positional error of $\le$0.4$''$. This is coincident (to within the errors) with an earlier ATCA measurement \citep{caswell98} and with a Very Large Array measurement at 1665 MHz \citep{argon00}. Position estimates of the strongest methanol features at 6.7 and 12.2 GHz \citep{caswell09a,xu11} are 18$^{\rm h}$11$^{\rm m}$51.40$^{\rm s}$, $-$17$^{\circ}$31$'$29.6$''$ and 18$^{\rm h}$11$^{\rm m}$51.396$^{\rm s}$, $-$17$^{\circ}$31$'$29.91$''$ respectively, a nominal offset from the hydroxyl of 0.7$''$$\pm$0.6$''$. At an astrometric distance of 2.3$\pm$0.1 kpc \citep{xu11}, this small angular offset would correspond to a physical offset of 1500 AU ($\sim$10 light days). We found that the peak flux densities of the main features in both the 1665-MHz and 1667-MHz transitions showed variability, but by less than 20$\pm$3\% from the maximum peak flux density with the exception of one epoch in the 1665-MHz transition and one epoch in the 1667-MHz transition, where emission dropped by 31$\pm$3\% and 38$\pm$3\% respectively. 
The averaged spectra and extrema for the ground state transitions are given in Figure \ref{rangefig1}. The average spectrum is the combination of the median value at each spectral channel taken from the values of all epochs of observations. Similarly the extrema are the highest and lowest values for a spectral channel across all epochs, excluding the furthest outliers (to avoid bias by any extreme noise fluctuations). 
%START OF NEW MATERIAL
The Gaussian fits used to construct the time series had errors of 0.01\,km\,s$^{-1}$ for both of the 1665-MHz features (found to be centred at 33.01\,km\,s$^{-1}$ and 35.26\,km\,s$^{-1}$) and  $<$0.01\,km\,s$^{-1}$, 0.02\,km\,s$^{-1}$, 0.02\,km\,s$^{-1}$ and 0.03\,km\,s$^{-1}$ for the four (weaker) 1667 MHz features (at 33.46\,km\,s$^{-1}$, 32.82\,km\,s$^{-1}$, 34.79\,km\,s$^{-1}$ and 31.53\,km\,s$^{-1}$ respectively). Any variation in the fitted peak velocity was within the errors and no systematic shift in the peak velocity of features was found. There was also no consistent shift in velocity for the epochs which had minima in the flux density.
%END OF NEW MATERIAL
The time series for the main features at each frequency are shown in Figures \ref{timeseriesfigure1} and \ref{timeseriesfigure2}. No emission was detected from the satellite line transitions at any of the epochs (to $\sim$0.2 Jy 5$\sigma$ limit). Additionally, for the last eight epochs the excited state transitions at 6030 MHz and 6035 MHz were observed, but also did not have any detected emission (to $\sim$0.2 Jy 5$\sigma$ limit).  

\begin{figure}
\begin{center}
\renewcommand{\baselinestretch}{1.1}
\includegraphics[width=7.5cm]{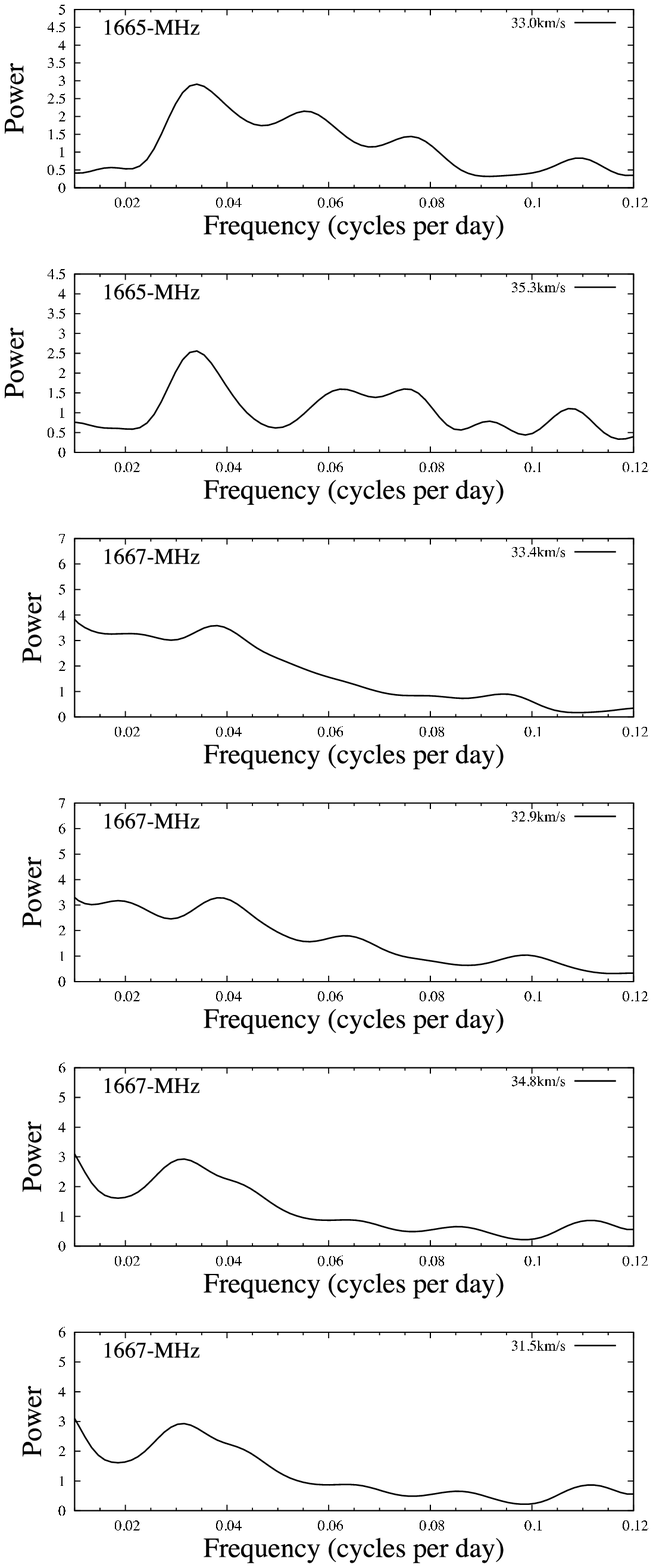} 
\caption{\small Results of Lomb-Scargle periodograms for the two 1665 MHz features and four 1667 MHz features.
The power spectra were oversampled by a factor of five and have been smoothed with a bezier function for clarity.
There is a suggestion of a peak at 0.03 to 0.04 cycles per day, which would correspond to a period of 25 to 30 days, coincident with the period of the 6.7-GHz methanol maser.}
\label{LombFig}
\end{center}
\end{figure}

%START OF NEW MATERIAL
A Lomb-Scargle periodogram \citep[following the algorithm of][]{press89} was used to search for any periodicity in each of the six features shown in the time series, with results displayed in Figure\,\ref{LombFig}.  At low significance there is a suggestion of a 0.03 to 0.04 cycles per day peak which would correspond to a period of 25 to 30 days, coincident with the period of the 6.7-GHz methanol maser. It is most noticeable in the 33.0 and 35.3\,km\,s$^{-1}$ peak features of the 1665-MHz transition which have power peaks at 0.034 cycles per day (29.4 days). This power peak was not affected by excluding the two highest flux density epochs from the periodogram. In contrast, after exclusion of the two lowest flux density epochs, the power peak was less evident for the 33\,km\,s$^{-1}$ feature and absent for the other features. This implies, comparable to the methanol, that the suggested periodicity is predominantly in the minima of the emission.
%END OF NEW MATERIAL

\begin{figure}
\begin{center}
\renewcommand{\baselinestretch}{1.1}
\includegraphics[width=8.0cm]{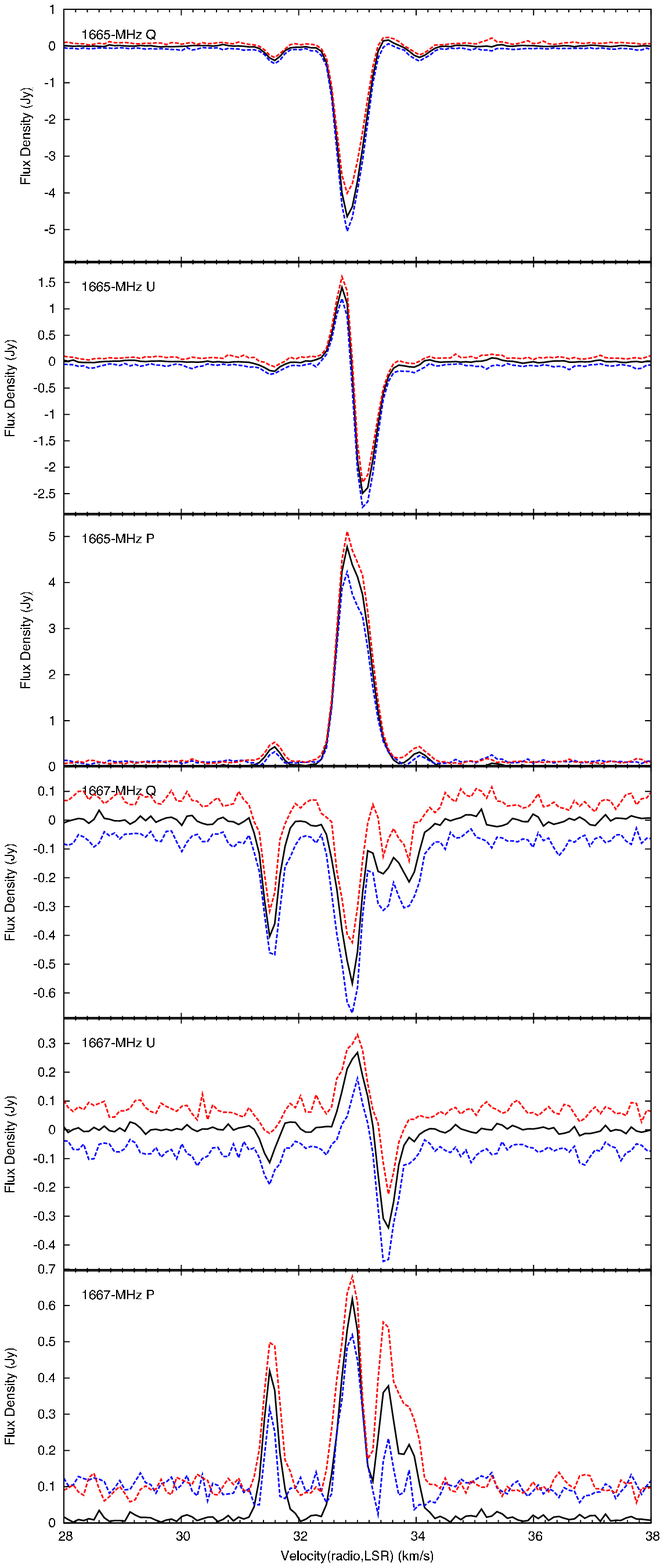} 
\caption{\small Range in flux density variation in Stokes $Q$ and $U$ and Linear Polarisation ($P$) for G12.889+0.489 at the 1665 and 1667 MHz hydroxyl transitions. The solid line is the averaged spectrum, constructed from the median value from all epochs for each spectral channel. The two dashed lines show the net extreme values (having excluded the most extreme noise-biased values).}
\label{rangefig2}
\end{center}
\end{figure}

\begin{figure}
\begin{center}
\renewcommand{\baselinestretch}{1.1}
\includegraphics[width=8.0cm]{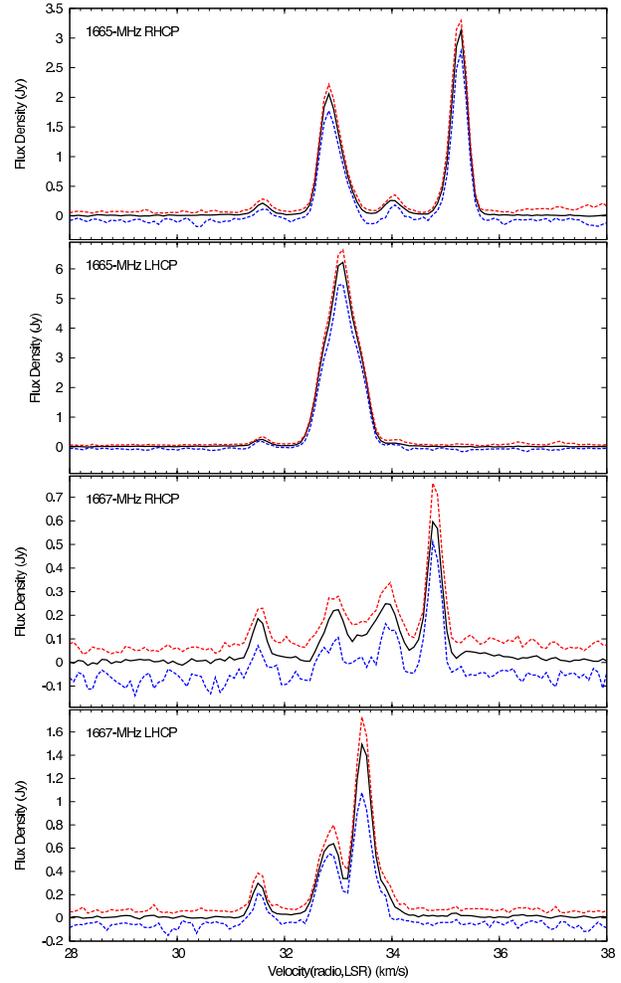} 
\caption{\small Range in flux density variation in RHCP and LHCP for G12.889+0.489 at the 1665 and 1667 MHz hydroxyl transitions. The solid line is the averaged spectrum, constructed from the median value from all epochs for each spectral channel. The two dashed lines show the net extreme values (having excluded the most extreme noise-biased values).}
\label{rangefig3}
\end{center}
\end{figure}

The features in the 1665-MHz Stokes $I$ spectrum at 31.6\,km\,s$^{-1}$, 33.0\,km\,s$^{-1}$ and 34.0\,km\,s$^{-1}$ were found on average to be 91\%, 55\% and 70\% linearly polarised respectively. The features in the 1667-MHz Stokes $I$ spectrum at 31.5\,km\,s$^{-1}$, 32.9\,km\,s$^{-1}$, 33.4\,km\,s$^{-1}$ and 34.8\,km\,s$^{-1}$ were found on average to be 84\%, 72\%, 23\% and 2\% linearly polarised respectively. The averaged and extreme spectra of the linear polarisation are shown in Figure \ref{rangefig2}. The right hand circular polarisation (RHCP) and left hand circular polarisation (LHCP) features, derived from RHCP = $(I+V)/2$ and LHCP = $(I-V)/2$,  are shown in Figure \ref{rangefig3}.
Spectra from \citet{argon00} taken in 1993 January, limited to 1665 MHz, and only RHCP and LHCP, show the same features as our spectra but typically at half the intensity.  Archival unpublished data from Parkes confirm the lower intensity near that epoch (1993 July).  The Argon et al. VLA data suggest that some spectral features are 
%NEW WORD:
spatial 
blends of several nearby components (although not well resolved by the VLA beam), but the separations of all major features in the velocity range 32 to 36\,km\,s$^{-1}$ are less than 0.15$''$.

In our monitoring sessions, of all the circularly polarised features, the RHCP features were found to vary by up to 33$\pm$3\% from the maximum and the LHCP features by up to 38$\pm$3\% from the maximum. Our measurement of the feature found at 31.6\,km\,s$^{-1}$ in both the 1665 and 1667 MHz transitions is of particular note. It is an outlying feature both in velocity and spatially. The 1665 MHz emission has an offset position of RA(J2000) 18$^{\rm h}$11$^{\rm m}$51.53$^{\rm s}$ and Dec(J2000) $-$17$^{\circ}$31$'$27.3$''$ (the 1667 MHz RA(J2000) 18$^{\rm h}$11$^{\rm m}$51.54$^{\rm s}$ and Dec(J2000) $-$17$^{\circ}$31$'$28.6$''$). Both transitions have high fractional linear polarisation. The polarisation position angle of this feature at 1665-MHz varies between 6$^{\circ}$ and 16$^{\circ}$ with a median angle of 12$^{\circ}$. The averaged spectrum has a position angle of 12$^{\circ}$. The position angle of this feature at 1667-MHz varies between 0$^{\circ}$ and 10$^{\circ}$ with a median angle of 5$^{\circ}$. The averaged spectrum has a position angle of 5$^{\circ}$. The error in the position angle due to noise is estimated to be 5$^{\circ}$.

\section{Discussion}
As stated in the introduction, the methanol maser counterpart of G12.889+0.489 has quasi-periodic flaring at both 6.7 GHz and 12.2 GHz \citep{goedhart09}. There are time delays between the two frequencies and the flaring period occurs anywhere within an 11 day window, but minima are regular (with a period of 29.5 days). This periodic variability was ascribed by the authors as likely being due either to variation in the radio continuum emission (the seed photons for the maser emission) or variation in the infrared emission (pumping the maser).  We note that, extrapolating from the work of \citet{goedhart09}, our range of epochs should have covered flaring activity and minima (three minima should occur in the first session of observations and the second session should be wholly between two adjacent minima). Goedhart et al. show that the methanol exhibits flux density variations of 60 to 70\%. With the exception of one epoch, the hydroxyl emission we measure varies by only 20\%. 
%following sentences varied
The current data have no strong indication of periodicity, but do suggest at low significance variation with a comparable period to the methanol, with the minimum at MJD 55379 nominally agreeing with the extrapolated minima epoch of the methanol. Further observing epochs will be required to make firm conclusions on the periodicity of variation.

The polarisation properties we observe are comparable to the work of \citet{szymczak09}, who measured the full polarisation properties at 1665 and 1667 MHz with all four Stokes parameters with the Nancay telescope in 2003. They found linear polarization of 87.3\% and 76.7\%  in two 1665-MHz features at 31.59 (the kinematic and spatial outlier mentioned previously) and 32.76 km\,s$^{-1}$ respectively, and both features had low circular polarisation at the epoch of the observations (14.6\% and 8.9\% respectively). If the persistently strong features, 1665-MHz RHCP at 35.3\,km\,s$^{-1}$ and the 1665-MHz LHCP at 33.1\,km\,s$^{-1}$, are a Zeeman pair, the implied magnetic field strength is +3.7 mG. This is corroborated with the 1667-MHz emission where the brightest RHCP at 34.8\,km\,s$^{-1}$ and the brightest LHCP feature at 33.4\,km\,s$^{-1}$, also give an implied magnetic field strength of +3.7 mG.

A qualitative explanation for some of the methanol maser sources with long periods (e.g. 9.62+0.20 and 188.95+0.89) is that of a colliding wind binary system \citep{walt09,walt11}, a system which could periodically alter either the seed flux or pumping mechanism of the maser emission (although for the example sources in the work of van der Walt et al. the seed flux is the more likely in view of their long periods). The hydrogen winds of the binary pre-main sequence stars either heat the circumstellar dust or cause additional ionisation of hydrogen surrounding the forming high mass star. 
In the case of G12.889+0.489, an upper limit to continuum emission of 1 mJy has been established \citep{walsh98}, but if there is a very weak region of ionised hydrogen providing the seed photons, the possible offset location of the hydroxyl relative to the methanol (by $\sim$1500 AU) may place the hydroxyl sufficiently far away so as to be minimally affected, explaining the lower prominence of periodicity in the hydroxyl emission. On the other hand the optical depth at the lower frequency of the hydroxyl emission could be such that the seed radiation does not respond as it does at the methanol frequency. An aspect of this model is that we may expect to see rotation of polarisation angle of features with the passing of shocks associated with the colliding winds (and the flares of maser emission). As mentioned in the results we have one feature at 31.6\,km\,s$^{-1}$ with high linear polarisation, but this does not show significant variation in polarisation angle (variation is within the noise).  

An alternative theory to account for periodicity in masers has been put forward by \citet{araya10}. This was proposed to explain the periodicity in formaldehyde and methanol maser emission in IRAS 18566+0408 through circumbinary disk accretion. In this model the accreting material heats the dust, increasing the photons pumping the maser. In this variety of model, an offset location could also be sufficient to diminish the effect on the hydroxyl emission. 

An important aspect of G12.889+0.489 is the shortness of the periodicity (29.5 days), which any model for the system must account for. Additionally, the suggestion that the minima of hydroxyl emission may coincide with the minima of the methanol implies that the periodicity may arise from a quenching or suppression mechanism, rather than flaring. Continued monitoring of this maser source across the various maser transitions is required to provide further insight.

\section{Summary} 
We have obtained the first monitoring observations of G12.889+0.489 with the Australia Telescope Compact Array, finding variability at the level of 20\% on timescales of days.
Despite no distinct flaring characteristics, the hydroxyl variability showed possible periodicity similar to that of methanol, with the minimum of emission at the same phase.
It will require a full programme of sensitive daily monitoring of G12.889+0.489 at the ground-state hydroxyl transitions to fully test the apparent periodic behaviour in the variability, and thus assess whether  the mechanism causing periodicity in the methanol maser counterpart is shared with the hydroxyl. 

\section*{Acknowledgments}
The authors thank the referee, Sharmila Goedhart, for useful comments which improved the paper. The authors also thank D.H.F.M.~Schnitzeler for observing assistance and the staff of the Australia Telescope Compact Array. The Australia Telescope Compact Array is part of the Australia Telescope which is funded by the Commonwealth of Australia for operation as a National Facility managed by CSIRO

\bibliographystyle{mn2e} \bibliography{UberRef}

\label{lastpage}

\end{document}